\documentclass[amsmath,amssymb,superscriptaddress,showpacs,twocolumn]{revtex4-1}
\usepackage{graphicx}

\begin{document}

\title{The colon-pile}

\author{Alexei Vazquez}
\email{alexei.vazquez.2@gmail.com}
\affiliation{Cancer Research UK Beatson Institute, Glasgow, United Kingdom}
\affiliation{Institute for Cancer Sciences, University of Glasgow, Glasgow, United Kingdom}

\date{\today}

\begin{abstract}
Bacteria populate the colon where they replicate and migrate in response to nutrient availability. Here I model the colon bacterial population as a sandpile model, the colon-pile. Sand addition mimics bacterial replication and grains toppling represents bacterial migration coupled to high population density. The numerical simulations reveal a behaviour similar to non-conservative sandpile models, approaching a critical state with system wide avalanches when the death rate becomes negligible. The critical exponents estimation indicates that the colon-pile belongs to a new universality class. This work suggest that the colon microbiome is in a self-organised critical state, where small perturbations can trigger large scale rearrangements, covering an area comparable to the system size and characterised by a $1/f$ noise spectra.
\end{abstract}

\maketitle


The bacterial population of the colon, the micobiome, is an ecosystem of several species linked through symbiotic relationships among themselves and with the host \cite{buchanan18}. The literature on colonic bacterial populations has focused on how  the host and environmental factors affect the diversity of bacterial species. Less attention has been paid to what is the overall bacterial load in normal physiological conditions. There are studies investigating the impact of antibiotics, which can lead to a large reduction in the gut bacterial population size  \cite{caporaso11}. Here I will instead focus on the natural variations of the number of colonic bacteria due to bacterial replication, migration and a small but finite death rate.

\begin{figure}
\includegraphics[width=3in]{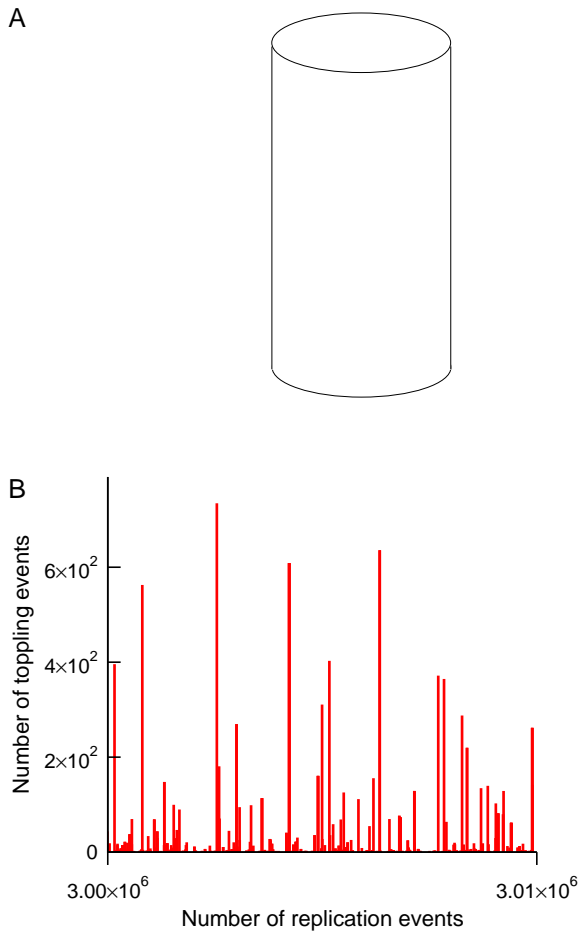}
\caption{a) Colon-pile geometry. b) Avalanches following bacterial replication events for a colon-pile of $h_c=4$ and $\epsilon=10^{-6}$.}
\label{fig1}
\end{figure}


I will model the colon as a tube (Fig. \ref{fig1}a). Basically a two dimensional grid with periodic boundary conditions along the tube circumference and closed boundary conditions at the tube ends. At each point of this lattice there will be a certain number of bacteria $h(x,y)$, the local height of the pile, where $x$ and $y$ are the coordinates on the tube. The anaerobic nature of the colon environment together with the complexity of the polysaccharides feeding the colonic bacteria makes their replication very slow. Based on this evidence I will assume that the bacterial replication takes place at an infinitesimal small rate. 

I will also take into account that bacterial migration is coupled to nutrient availability \cite{ni16}. When the local nutrient density is high, bacteria will tend to allocate their metabolic resources to fuel their replication. In contrast, when nutrients are scarce, bacteria will tend to allocate their metabolic resources to migrate in search for nutrients. When the bacterial number at a given location is too high, we expect a depletion of the nutrient concentration at that location. This local nutrient depletion will then induce the migration or death of bacteria at the corresponding coordinates. I will put all these elements together into a threshold dependent migration rule. When the local bacterial count $h(x,y)$ exceeds a threshold $h_c$, each bacteria at that position dies with probability $\epsilon$ or otherwise migrates to one of the nearest neighbours locations. I will refer to this event as a toppling, following the language of sandpile models.


This model is quite similar to the dissipative sandpile models that were investigated in the context of self-organised criticality \cite{chessa98,vazquez00}. Bacterial replication plays the role of sand addition and bacterial migration coupled to population density mimics sand grains toppling. Following the language of sandpile models, the cascade of toppling events associated with the addition of one grain is called an avalanche.The number of toppling events is called the avalanche size and it is denoted by $S$.  Based on experience from dissipative sandpile models \cite{chessa98,vazquez00}, the system approaches a self-organised critical state when $\epsilon\rightarrow0$.  Near the critical state the avalanche size and other avalanche characteristics follow a power law distribution, one signature of self-organised criticality \cite{bak87}. More precisely, the distribution of avalanche sizes and other avalanche characteristics satisfy the scaling form
\begin{equation}
P(x) = x^{-\tau_s} f(x/x_c)
\label{distribution}
\end{equation}
\begin{equation}
x_c \sim \epsilon^{-d_x}
\label{cutoff}
\end{equation}
where $\tau_x$ and $d_x$ are scaling exponents and $x=S$ or other avalanche parameter.

Particle addition is however different than in dissipative sandpile models. In the colon-pile the addition of new grains is coupled to the local particle density as a birth process. The addition of the next grain will happen with higher probability at the coordinates that already have more particles. The question I address next is wether that changes the universality class.

\begin{figure}[t]
\includegraphics[width=3in]{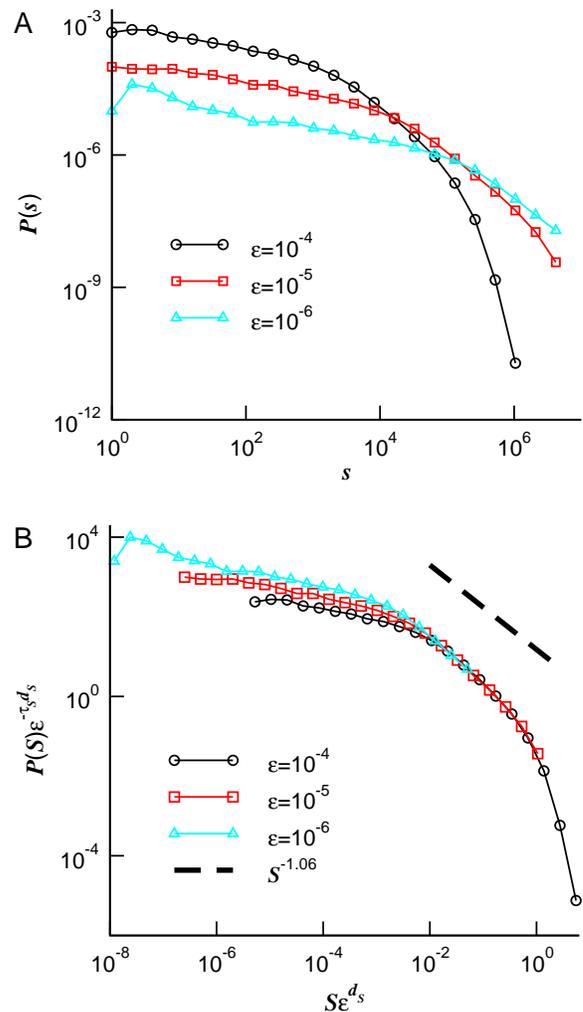}
\caption{a) Avalanche size distribution for $h_c=4$ and different values of $\epsilon$, using logarithmic binning. b) After rescaling.}
\label{fig2}
\end{figure}

\begin{figure}[t]
\includegraphics[width=3in]{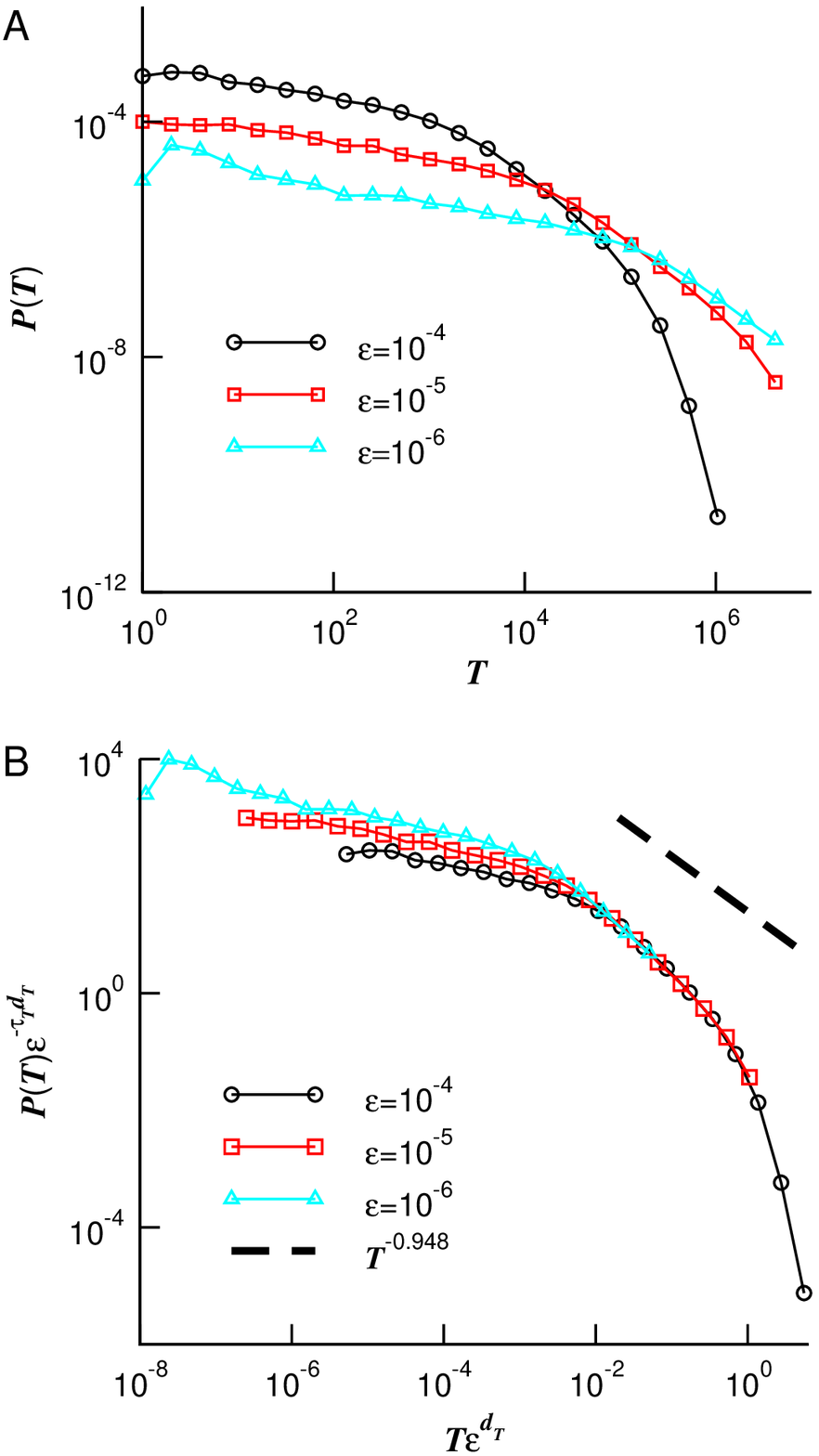}
\caption{a) Avalanche duration distribution for $h_c=4$ and different values of $\epsilon$, using logarithmic binning. b) After rescaling.}
\label{fig3}
\end{figure}

\begin{figure}[h]
\includegraphics[width=3in]{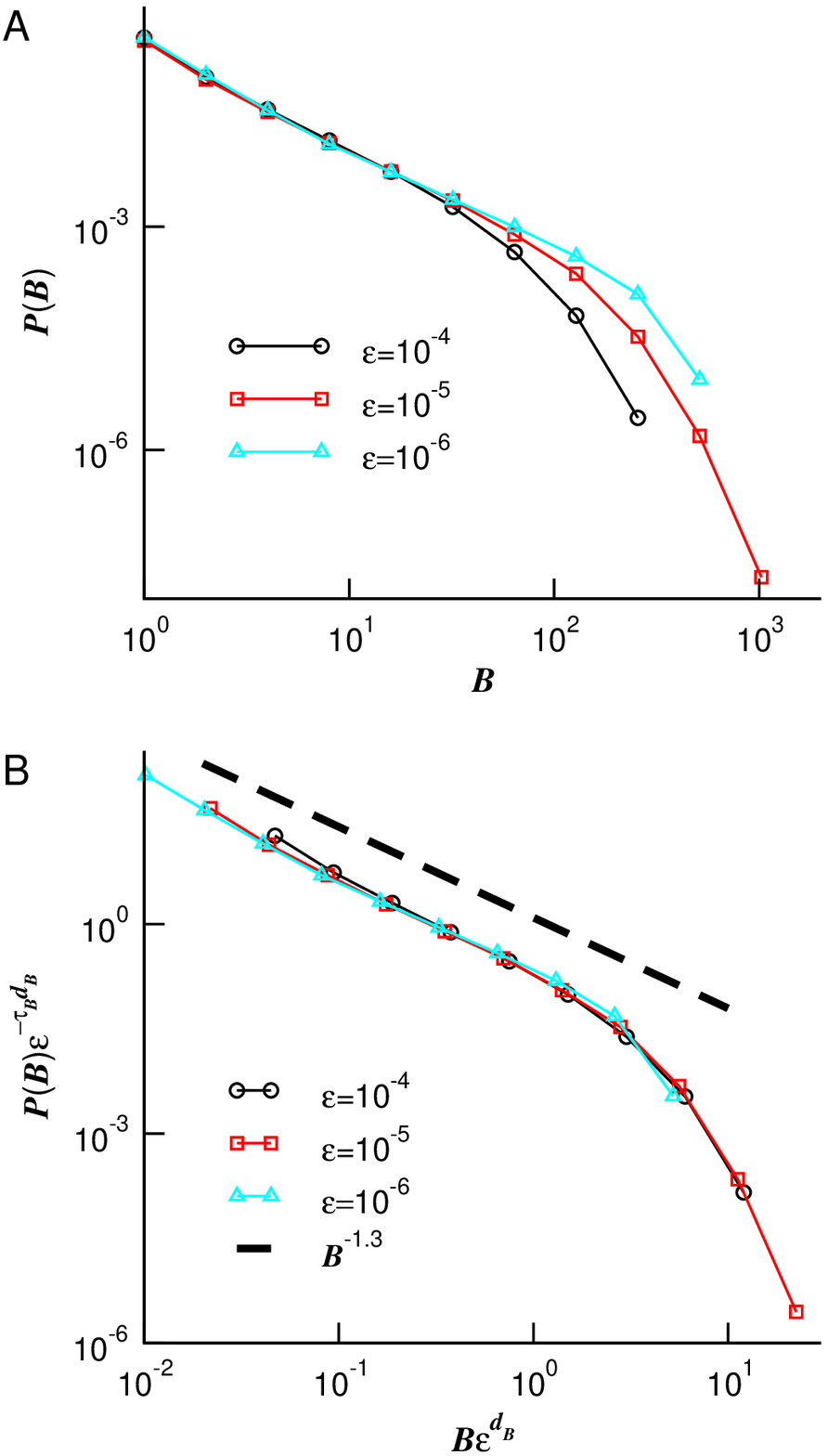}
\caption{a) Avalanche burden distribution for $h_c=4$ and different values of $\epsilon$, using logarithmic binning. b) After rescaling.}
\label{fig4}
\end{figure}

\begin{figure}[h]
\includegraphics[width=3in]{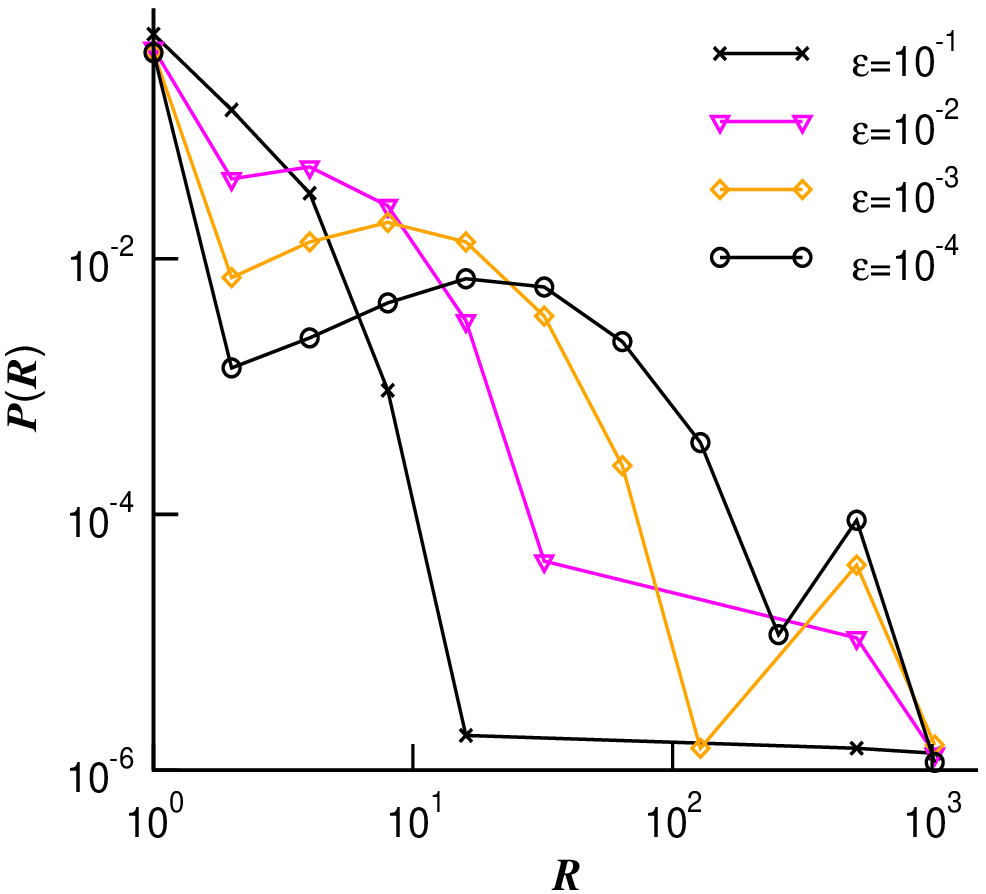}
\caption{a) Avalanche radius distribution for $h_c=4$ and different values of $\epsilon$, using logarithmic binning. b) After rescaling.}
\label{fig5}
\end{figure}


I will use numerical simulations to estimate the critical exponents of the colon-pile model. The dynamics is divided into particle addition and avalanche dynamics. I create a particle list containing all the particles (bacteria) in the colon-pile and a two-dimensional array keeping track of the colon-pile local heights $h(x,y)$. {\em Particle addition:} At each particle addition step, I select a particle from the list with equal probability, add a duplicate of it to the list and update the pile height at the associated position $h(x,y)\rightarrow h(x,y)+1$.  {\em Toppling:} If $h(x,y)\geq h_c$ at any position in the pile, then each particle at that position is removed with probability $\epsilon$ or otherwise moved to one of the neighbour positions $(x-1,y)$, $(x,y-1)$, $(x+1,y)$ or $(x,y+1)$ with equal probability. This update is done synchronously for all current sites with $h(x,y)\geq h_c$.

I have carried on numerical simulations for a tube of dimension $1024\times1024$, dissipation parameter values $\epsilon=10^{-1},\cdots,10^{-6}$ and local heigh thresholds $h_c=4,8,16$. The colon-pile was initialised by assigning a high in the range $[0,\ldots,h_c)$ with equal probability. To avoid the initial transient states, I ran the model for 100,000 avalanches before recording . Then I recorded 100,000 avalanches to carry on the statistics.   

Figure \ref{fig1}b illustrates the type of dynamics generated by the colon-pile, using the number of replication events as a clock.The number of toppling events after a single bacterial replication follows a wide distribution, with several small and frequent large spikes in the number of toppling events. The avalanche size distribution $P(S)$ follows a wide distribution with a cutoff for large sizes (Fig. \ref{fig2}a). The cutoff shifts to the right with decreasing $\epsilon$, corroborating the approach to a critical state.

Based on the scaling form (\ref{distribution}), the distribution moments satisfy the scaling
\begin{equation}
\left<x^n\right> = \int ds P(x)x^n \sim \epsilon^{-q_{x,n}}
\label{moment}
\end{equation}
\begin{equation}
q_{x,n} = (1-\tau_x)d_x + n d_x 
\label{moment_exponent}
\end{equation}
where $x=S$. First I estimated $q_{x,n}$ from a linear fit to the plot of $\log\left<x^n\right>$ versus $\log\epsilon$, for $n=1,2,3,4,5$. Then I estimated $\tau_x$ and $d_x$ from a linear fit to the plot of $q_{x,n}$ versus $n$. The results are reported in Table \ref{exponents} for the avalanche size and other properties discussed below.

\begin{table}[t]
\begin{tabular}{l|l|l|l|l|l|l}
$h_c$ & $\tau_S$ & $d_S$ & $\tau_T$ & $d_T$ & $\tau_B$ & $d_B$\\
\hline
4 & 1.06 & 1.32 & 1.03 & 0.266 & 1.30 & 0.332 \\
8 & 1.07 & 1.32 & 0.948 & 0.261  & 1.31 & 0.326 \\
16 & 1.07  & 1.31 & 0.938 & 0.271 & 1.32 & 0.316 \\
\hline
\end{tabular}
\caption{Scaling exponents}
\label{exponents}
\end{table}

The rescaling of the avalanche size distribution, using the scaling exponents in Table \ref{exponents}, exhibits an overlap for intermediate and large values of the horizontal axis. The overlap breaks down for small sizes, indicating that $P(S)$ cannot be reduced to the scaling form (\ref{distribution}) with one scaling parameter. The scaling deviations for small sizes are related to the power law exponent $\tau_S$ being close to 1. In such a case the full distribution shifts to lower values in the vertical axis as the tail shifts to the right (Fig. \ref{fig2}a).

The estimated power law exponent $\tau_S=1.06-1.07$ is different than the 1.25 reported by Chessa et al \cite{chessa98} and the 1.1 reported by Vazquez \cite{vazquez00} for two different variants of the dissipative sandpile model. Furthermore, the avalanche size distributions for those dissipative sandpile models follow the simple scaling form (\ref{distribution}) across the whole range of avalanche sizes, indicating that this is not just a matter of the exponents accuracy. Therefore, modelling the particle addition as a birth process puts the colon-pile in a new universality class.

There are other interesting properties of the colon-pile with biological relevance. The statistics of the inter-avalanche time, here denoted by $T$,  can give us an idea of whether the colon-pile follows a standard Poisson dynamics with an exponential inter-event time distribution, or punctuated equilibria with a wide distribution of inter-vent times. The inter-avalanche times follow a wide distribution $P(T)$ that behaves similar to the avalanche size distribution when $\epsilon\rightarrow0$ (Fig. \ref{fig3}a). $P(T)$ can be rescaled for intermediate and large inter-event times but the scaling breaks down for small inter-event times (Fig. \ref{fig3}b). The frequency spectrum associated with the avalanche events is given by \cite{bak87}
\begin{equation}
{\cal S}(\omega) = \int dT \frac{P(T)T}{1+(\omega T)^2} \sim \omega^{-2+\tau_T}
\label{spectrum}
\end{equation}
Since $\tau_T\approx1$ the avalanche time series exhibits a $1/f$ noise spectra. 

Another biologically relevant quantity is how many bacteria die during an avalanche. The number of dissipated particles, the dying bacteria, will be called the avalanche burden and it will be denoted by $B$. The burden size follows a broad distribution $P(B)$ with a cutoff that shifts to the right when $\epsilon\rightarrow0$ (Fig. \ref{fig4}a). $P(B)$ follows the simple scaling form (\ref{distribution}) across all values, as demonstrated in Fig. \ref{fig4}b. The colon-pile is thus characterised by frequent loss of a small number of bacteria and less frequent but probable events where a large number of bacteria die. In theory, these large death events could lead to the extinction of subpopulations of bacterial species.

Finally, we are also interested in what is the typical spatial area associated with an avalanche. I will define the avalanche radius, denoted by $R$, as the maximum distance from the first site of replication to any site perturbed by the avalanche, either along the circumference or the longitudinal axes. The estimated distribution $P(R)$ extends from 1 to the system size already starting from $\epsilon=0.01$ (Fig. \ref{fig5}). The shape of the distribution does not follow the scaling form (\ref{distribution}) but it is characterised by an increase in the frequency of avalanches with large radius when $\epsilon\rightarrow0$. 


The colon-pile model indicates that the gut  bacterial population is in a self-organised critical state. This state is characterised by large-scale rearrangements in the bacterial population numbers following the replication of a single bacteria. The magnitude an non-locality of these reorganisation of the colon-pile are more extreme the smaller is the death rate associated with nutrient scarcity. The validity of these theoretical observations remains to be determined experimentally.

From the theoretical point of view, the critical state is a consequence of the coupling between migration and population density. This coupling is encoded in the topping rule, migration happens when the population density exceeds a threshold. It is precisely this toppling rule what brings in the analogy with the sandpile model, the prototype of self-organised criticality \cite{bak87}. The key difference with the canonical sandpile models is the particle addition. In the colon-pile particle addition is encoded as a birth process. Based on the numerical estimates this difference puts the colon-pile in a new universality class.

\bibliographystyle{unsrt}
\bibliography{ref_sandpile.bib}

\end{document}